\documentclass[prb, twocolumn, letter, superscriptaddress]{revtex4}
\usepackage{amssymb}
\usepackage{amsmath}
\usepackage{bbm}
\usepackage{graphicx}
\usepackage{color}

\begin{document}

\title{The Interplay of Nonlinearity and Architecture in Equilibrium Cytoskeletal Mechanics}

\author{Shenshen Wang}
\affiliation{Department of Physics, Department of Chemistry and Biochemistry,  and Center for Theoretical Biological Physics, University of California, San Diego, La Jolla, CA 92093, USA}
\author{Tongye Shen}
\affiliation{Department of Biochemistry, Cellular and Molecular Biology,
University of Tennessee, Knoxville, TN 37996, USA}
\author{Peter G. Wolynes}
\affiliation{Department of Physics, Department of Chemistry and Biochemistry,  and Center for Theoretical Biological Physics, University of California, San Diego, La Jolla, CA 92093, USA}

\date{\today}

\begin{abstract}

The interplay between cytoskeletal architecture and the nonlinearity of the interactions due to bucklable filaments plays a key role in modulating the cell's mechanical stability and affecting its structural rearrangements. We study a model of cytoskeletal structure treating it as an amorphous network of hard centers rigidly cross-linked by nonlinear elastic strings, neglecting the effects of motorization. Using simulations along with a self-consistent phonon method, we show that this minimal model exhibits diverse thermodynamically stable mechanical phases that depend on excluded volume, crosslink concentration, filament length and stiffness. Within the framework set by the free energy functional formulation and making use of the random first order transition theory of structural glasses, we further estimate the characteristic densities for a kinetic glass transition to occur in this model system. Network connectivity strongly modulates the transition boundaries between various equilibrium phases, as well as the kinetic glass transition density.

\end{abstract}


\maketitle

\section{Introduction}

The cytoskeleton is a crowded network of dynamic filamentous protein polymers that collaborate with diverse binding proteins and molecular motors to form nature's most marvelous active material \cite{cytoskeleton}.
All eukaryotic cells are known to contain a well-developed cytoskeleton, and even bacteria have been found to contain a diverse set of proteins capable of forming structural filaments \cite{bacterial CSK}.
Actin is a major constituent of the cytoskeletal network that participates in such widely differing processes as blood clotting, egg fertilization, intestinal absorption and tumor invasion. A common theme of many substantial experimental efforts \cite{actin mechanics, assem_disassem, biophy para, loading history, stress softening} is the role of actin filaments in maintaining cell architecture and generating movement.

\textit{In vitro}, actin can polymerize to form long rigid filaments (F-actin), with a diameter around 7 $\mathrm {nm}$ and contour length up to 20 $\mathrm {\mu m}$. The \textit{in vivo} cytoskeletal network, however, is regulated and controlled not only by the concentration of F-actin, but also by accessory proteins that bind to F-actin. Nature provides a host of actin-binding proteins (ABPs) with versatile functions that offer the necessary variations for the part actin has to play \cite{ABP}: Cross-linking proteins form filament bundles or isotropic gels, whereas capping proteins and/or filament-severing proteins regulate actin polymerization under specific salt conditions and thus control the average length of F-actin. Experiments show, for example, that actin-binding protein functions to stabilize cortical actin \textit{in vivo} and is required for efficient cell locomotion \cite{cortical stability}. Rheological experiments on simplified networks and living cells demonstrate that even the most basic mechanical properties of cytoskeletal materials are sensitive to specific architectural details, including filament and cross-link density, connectivity, and orientation \cite {mechanical properties}.

Unlike flexible polymers, where changes in cross-link density typically do not markedly affect the elasticity, small changes in the concentration of cross-linking actin-binding proteins do dramatically alter the elasticity of F-actin networks \cite{semiflexible polymer}. As materials, \textit{in vitro} networks of cytoskeletal filaments exhibit several unusual mechanical properties, including a highly nonlinear elastic response \cite{stress softening, nonlinear response} and negative normal stresses \cite{negative normal stress}.
Another noteworthy observation is the buckling of actin stress fibers that occurs upon rapid shortening \cite{actin buckling}, which demonstrates the instability of a prestressed actin network under compression.


A refined experimental route to understanding cytoskeletal structures in terms of their molecular components is to reconstitute these structures from purified proteins. \textit{In vitro} reconstitution provides the capability to study the emergence of micrometer-scale function from numerous molecular-scale interactions. A recent reconstitution of contractility in a simplified model system, composed of purified F-actin, muscle myosin II motors and $\alpha$-actinin cross-linkers, has shown that contractility occurs above a threshold motor concentration (one myosin filament for every $30$ actin filaments) and within a window of cross-link concentrations ($90-270$ $\alpha$-actinin dimers per actin filament) \cite{concentration window}. It is somewhat unexpected that at high cross-link densities the bundled networks do not contract on the experimental timescale of an hour. This seems to imply a dramatic slow-down of cytoskeletal kinetics due to the steric constraints associated with increasing crowding.



The cytoskeleton in different organisms or even during different stages of the cell cycle exhibits a rich variety of viscoelastic properties \cite {JJFredberg1, JJFredberg2}. To accommodate both the adaptive behavior of dynamic remodeling, and the ability to stabilize and be resistant to deformation, the cytoskeleton should exhibit at least two mechanical phases: a plastic/fluid phase for internal reorganization of the cell, and an elastic/solid phase for mechanical support and tension transmission.

In contrast to the strain-stiffening behavior \cite{strain stiffening} in response to a \textit{sustained} stretching of cells or reconstituted cross-linked actin gels, measurements on the responses to a \textit{transient} stretching in the living cells \cite {transient stretch} demonstrate that the cell will fluidize in response to the stretch, but will resolidify subsequent to this fluidization. This ability seems to be insensitive to molecular details, and instead only depends on the proximity of the thermodynamic conditions of the cell to a solid-like state before the stretch. This observation implied that in addition to specific signalling pathways, some response mechanisms are most likely controlled by non-specific actions of a slowly evolving network of physical forces.
On the other hand, the universal behavior observed in the osmotically compressed cell \cite{osmotically compressed cell} highlighted the crowding-induced stiffening of the cytoplasm as the solid volume fraction is sufficiently high, and suggested an analogy to the colloidal glass transition.

In an attempt to understand the interplay between the individual filament properties and network architecture, we study in this paper the equilibrium properties of a model cytoskeleton as an amorphous network of rigidly cross-linked nonlinear strings which also contains nodes with excluded volume. We incorporate the nonlinearity of interaction due to bucklable filaments into the force law, and characterize the cytoskeletal architecture by cross-link density and network connectivity. A mean-field level investigation, within the framework of the self-consistent phonon method and a density functional formulation, reveals a diversity of mechanical phases and a number of possible transitions in between which can be controlled by biophysical parameters. This phase diagram may shed light on our understanding of cytoskeletal remodeling in response to mechanical or chemical stimuli. We also show the possibility of a glass transition in this model system and how the network connectivity modulates the transition densities.


\section{{Model and method}\label{methods}}
\subsection{Model}
\begin{figure}[htb]
\centerline{\includegraphics[angle=0, scale=0.28]{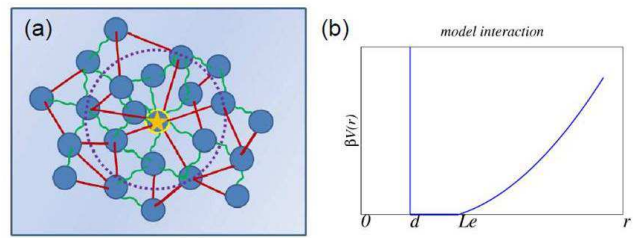}}
\caption{Illustrations of the model system and the nonlinear interaction.
(a) The beads (blue spheres) interconnect the F-actin (red straight or green squiggly lines) into an amorphous network. Red straight lines stand for tense/stretched bonds and green squiggly lines for loose/buckled bonds. The arbitrarily chosen central particle (yellow-stared) is connected with its nearest neighbors within the first shell of the radial distribution function (the dotted purple circle).
(b) Nearest-neighbor interaction versus radial separation.}
\label{interaction_model}
\end{figure}

We model the F-actin bundles as individual nonlinear elastic strings; they are capable of resisting tensile forces by stretching (beyond the relaxed length $L_e$), but are unstable under compression so that these forces may cause the strings to buckle. The actin-binding proteins bundle and crosslink F-actin to form an amorphous network. To account for the excluded volume of F-actin and the ABP aggregates, to the lowest order, we model the network as being crosslinked by hard ``beads"; each bead serves as a compact rigid subunit that concentrates the volume of the F-actin and ABP aggregates centered on that bead. The network elasticity then comes from the remaining inter-bead F-actin molecules which are now taken to be volumeless. To higher order, the elasticity within the beads can also be included by introducing softness in the repulsion. In this way we effectively decompose the inter-bead interaction into purely excluded volume and nonlinear elastic contributions, and the model potential energy $V(r)$ between a nearest-neighbor pair of beads is given by
\begin{equation}
\beta V(r)=A\,\Theta(d-r)+\frac{\beta\gamma}{2}(r-L_e)^2\,\Theta(r-L_e)\,,
\end{equation}
where $\Theta(x)$ is the Heaviside step function. The limit $A\rightarrow\infty$ indicates the hard-sphere (HS) repulsion, while $\gamma$ measures the rigidity of the inter-bead F-actin. Temperature dependence enters this model only via the combination $\beta\gamma$ where $\beta=1/k_B T$. Here $d$ denotes the HS diameter of the beads and $L_e$ marks the onset of elasticity.
Thus $L_e>d$ define a buckling regime ($d<r<L_e$) where no load is imposed.
We show a sketch of the model system (Fig.~\ref{interaction_model}a) and a schematic of the nonlinear interaction (Fig.~\ref{interaction_model}b).

We incorporate the filament nonlinearity to the extent that it effectively encodes the asymmetric response of the filaments to stretch and to compression; in this sense, our model is more ``coarse-grained" than the well-studied ``semi-flexible fiber" model \cite{semi-flexible fiber}; the latter treats the bending degrees of freedom of F-actin based on the contour of the filaments using a continuum description. This bending effect, stated in our language of inter-bead interaction, essentially introduces a finite resistance/restoring force to compression in our originally buckling regime. It can be easily incorporated, yet we do not expect any qualitative modification on the mean-field level predictions made by our current model.

Similar buckling bonds were used to study the statistical mechanics of a ``cat's cradle" built on a regular lattice by two of us several years ago \cite{cat's craddle}. Here we also include a HS repulsion to account for the excluded volume of the F-actin and ABP aggregates, and also assume the cytoskeleton adopts an aperiodic amorphous structure characteristic of the disorder of biological cytoskeletal networks. Thus in our model system, localization is achieved not only at a high concentration of beads via hard-core repulsions\,(topological caging), but also will occur upon network expansion due to bond stretching. The interplay between the nonlinearity of interaction due to bucklable bonds and the network architecture renders the thermodynamic state diagram nontrivial in terms of diverse mechanical phases. These equilibrium phases include persistent uniform liquid-like states, regions with a coexistence of frozen and liquid phases, and the possibility of a martensitic-like phase transition
that signals a spontaneous symmetry breaking.

The average length of F-actin in an \textit{in vitro} network can be adjusted by controlling
the concentration of capping proteins (for example the concentration of gelsolin); a higher molar ratio of capping protein to actin results in shorter F-actin on average.  Increasing the crosslink concentration promotes the formation of F-actin and ABP aggregates \cite{clumps}, and in turn increases the bead density. The bead density determines the average spacing between the beads, or equivalently, the average end-to-end distance of F-actin ($r$).
When $r<L_e$ the elasticity of the fiber is entropic in origin, whereas the intrinsic elastic modulus of the fiber dominates when $r>L_e$.
The effective stiffness of the interbead F-actin ($\beta\gamma$) is also variable; it is enhanced when filaments are
bundled together to form structures with larger diameters \cite{latch}.
Thermal fluctuations play a smaller role as filaments become stiffer.

To get a feeling about how the relevant biophysical parameters work out in our model system, we can first estimate the bead density $\rho$ from experimental actin concentration. The typical actin concentration used for \textit{in vitro} networks is $23.8\mu\mathrm{M}$ which, if we assume the actin monomers to be spheres of diameter $5\mathrm{nm}$, corresponds to a volume fraction of $0.9\times10^{-3}$. The ABPs take up negligibly small volume compared to that taken up by actin. We further take the bead size to be $1\mu\mathrm{m}$ and the F-actin as slender rods of cross section $5\mathrm{nm}\times5\mathrm{nm}$, then if each bead concentrates five F-actin of length $6\mu\mathrm{m}$,
$\rho=1.2$ is needed to reach the given volume fraction; if five F-actin of length $10\mu\mathrm{m}$ per bead
then $\rho=0.9$ (taking the bead diameter $d$ to be the length unit). Thus $\rho$ is adjustable, ranging from
$0.1$ to $1.4$, by varying actin concentration and/or crosslinking properties.
Since the filament aggregates are not perfectly dense-packed within the bead, they only take up a portion of the assumed bead volume; the modeled hard-sphere repulsion between the beads may overestimate the excluded volume effect. Yet the qualitative phase behavior of this model system should not be altered.
In a first approximation,
the cytoskeleton determines the mechanical properties of a cell; since the elastic modulus of a cell is in the range of $10^3$ Pa, the corresponding effective stiffness $\beta\gamma$ is then estimated to be between $1$ and $10$ as converted into our model parameters. (The characteristic bead size is $d=1\mu\mathrm{m}$.)


\subsection{Mean field approximations and the self consistent phonon (SCP) method}

Instead of treating the bending degrees of freedom of F-actin directly, we focus on the motion of individual beads,
located on the vertices of the network, in order to find the nonlinear elasticity as the bead density is varied.
In the mean-field spirit, we tag a given bead as the central particle,
and study its stability/response to the local mechanical environment.

We build the model system on top of an amorphous structure. In contrast to a
regular lattice that usually has a unique equilibrium configuration as well as a definite coordination number for a
specific lattice structure, amorphous systems must be appropriately averaged over non-vibrational disorder in the
lattice which we take as quenched. A further mean-field approximation will be made in the present analysis to avoid
detailing the configurational complexity of a random network. We will summarize the underlying
topology of the amorphous solid in an assumed isotropic pair distribution
function $g(r)$ for the fiducial configurations of the system. Such a treatment has also been used for molecular structural glasses. We then define nearest neighbors as those beads that sit within the
``first shell", i.e.,\,up to the first minimum of the equilibrium radial distribution function $g(r)$, and assume that interaction only exists between nearest-neighbor pairs.

\begin{figure}[htb]
\begin{center}
\includegraphics[angle=0, scale=0.26]{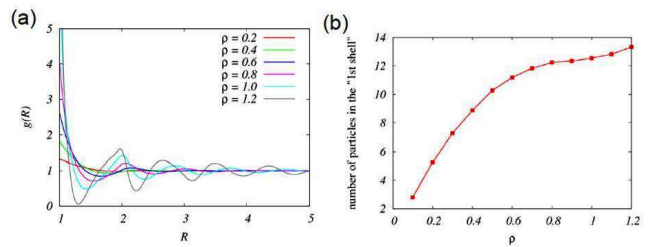}
\caption{(a)\,Radial distribution function $g(R)$ vs the radial separation $R$ for a series of particle density $\rho$. As density increases, oscillatory amplitude gets larger along with a phase shift toward the core. (b)\,Coordination number vs particle density.}
\label{radialDist_1stShell}
\end{center}
\end{figure}

For liquids above the melting point, $g_{HS}$ is
well described by the Percus-Yevick approximation; the Verlet-Weis
correction improves the behavior of $g(r)$ near the core and
dampens its oscillations at large $r$, giving accurate
modifications especially at high densities \cite{Verlet_Weis}.
We follow the procedures of Verlet and Weis, and note that
coordination number in an amorphous structure depends
on particle concentration, because the amplitude and phase of the oscillations in radial
distribution vary with number density/packing fraction of the particles (See Fig.~\ref{radialDist_1stShell}(a)).
As shown in Fig.~\ref{radialDist_1stShell}(b), an increasing number of nearest neighbors are
accommodated within the first shell when particles are packed
denser.
This increase is almost linear when $\rho$ is low, and slows down as $\rho$ approaches $0.6$; the modest increase above densities of $0.8$ is bounded by the random close packing value of the coordination number which is around $14$.

In the context of our model network, the amorphous topology effectively contains the physical aspects of bond breaking and/or ABP detachment upon network expansion. These aspects were absent in the lattice setting which was studied earlier: as the density of beads decreases (or equivalently as the network expands), coordination number drops off resulting in a smaller number of interacting neighbors or an effectively weaker network connectivity, even though the bond connections between each nearest-neighbor pair are assumed to be permanent at each given bead density.



As an approach of thermal stability analysis in an equilibrium
system, the SCP theory was first developed to treat the anharmonic effects
of hard-sphere crystals. The basic idea is to introduce a
reference harmonic system, and then obtain the effective potential felt by
each tagged particle, by averaging the pair interaction
over the assumed Gaussian fluctuations from all its neighbors.
This procedure should give back the assumed harmonic potential of the typical particle.
The resulting coupled set of self-consistent equations allows an
iterative scheme to determine the generally site-dependent force
constants. Among several schemes, Fixman's SCP method based on a
systematic expansion in Hermite functions has proven an efficient
and especially robust procedure \cite{fixman}. Recently, this technique has been
applied to network glasses \cite{network glasses}, and also by two of us to study motorized particle
assembly \cite{nonequil assembly} to analyze the
far-from-equilibrium dynamics like that of the living protoplasm using a local feedback scheme.

We apply the SCP method to quantify the responses of various mechanical
phases under varying physical conditions. At low concentration of beads,
the effective attraction due to stretched springs dominates,
thus the tagged particles are localized by``bond trapping";
whereas in the high concentration limit, HS repulsion dominates
and results in a glassy/jammed state owing to ``topological caging".
To investigate both localized phases and the intermediate states that bridge the transition,
we make a Gaussian local density profile ansatz with a single parameter, i.e. we describe the
time-averaged density configuration as a sum of Gaussians
representing thermal vibrations of particles about the fiducial sites
\begin{equation}
\rho(\vec{r})=\sum_i\left(\frac{\alpha_i}{\pi}\right)^{3/2}e^{-\alpha_i\left(\vec{r}-\vec{R}\right)^2}.
\end{equation}
In the present work, the force constants (inverse mean squared
displacements or localization strengths) $\{\alpha_i\}$ will all be
taken to be equal, but it is not difficult to allow spatial variation \cite{magnetic_analogy}, i.e.,\,dependence on the index i.

In the independent-oscillator version of the SCP theory, the effective
potential between two interacting particles is given by
\begin{equation}
e^{-\beta V^{eff}(|\vec{r}-\vec{R}'|;\,\alpha)}=
\left(\frac{\alpha}{\pi}\right)^{3/2}\int{d\vec{r}'e^{-\beta\frac{1}{2}V(\vec{r}-\vec{r}')}e^{-\alpha(\vec{r}'-\vec{R}')^2}},
\end{equation}
which may be explicitly written as
\begin{eqnarray}
e^{-\beta V^{eff}(R;\,\alpha)}&=&
\sqrt{\frac{\alpha}{\pi}}\frac{1}{R}\int_0^{\infty}dw
we^{-\frac{1}{2}\beta
V(w)}\nonumber\\
&\times&\left[e^{-\alpha(w-R)^2}-e^{-\alpha(w+R)^2}\right].
\end{eqnarray}
Here, $R$ denotes the averaged equilibrium separation between
interacting particles, and $\alpha$ represents the homogeneous
localization strength.

Taylor expansion of the effective interaction up to the second order
gives a self-consistent relation for $\alpha$:
\begin{equation} \label{alpha}
\alpha=\frac{\rho}{6}\int_{``1st\,
shell"}d^3\vec{R}\,g(\rho,R)\,Tr\!\left[\nabla\nabla\beta
V^{eff}(R,\alpha)\right],
\end{equation}
or
\begin{equation}
\alpha=\frac{2\pi}{3}\rho\int_1^{R^*}dR\,R^2g(\rho,R)\nabla^2\beta
V^{eff}(R,\alpha)\,.
\end{equation}
Here $R^*$ marks the position of the first minimum of the radial
distribution function $g(R)$. The HS diameter $d$, as the lower
limit, is taken to be the unit of length.

\subsection{Thermodynamic ramifications of our model system}

We shall show in a moment that in our model system---a rigidly cross-linked
nonlinear-elastic network---there exist at least two different localized
phases: one modestly localized phase describes the weakly trapped motion due to
bond stretching. We refer to this state as the ``liquid-like state", in view of
its considerable mobility and small localization strength
($\alpha_{liq}$). The other more strongly localized state corresponds to the jammed motion
within topological cages of neighbors. This solution depicts a ``glassy
state" exhibiting highly restricted vibrations ($\alpha_{gl}$).


Traditionally, microscopic treatments of liquids take the view
that since the liquid structure is dominated by repulsive forces,
it is desirable to develop perturbation theories based on a HS
reference system and then find the optimal parameters for it. In
this spirit, the Helmholtz free energy can be obtained by adding a
first-order perturbation to the free energy of the corresponding
HS system. On the other hand, the Carnahan-Starling equation of state
gives accurate values of the reference free energy at moderately
high densities. In sum, the free energy for the liquid-like state
is given by
\begin{eqnarray} \label{f_liq}
f_{liq}&\equiv&
\frac{\beta A_{liq}}{N}=\left(\ln\rho\Lambda^3-1\right)+\int_{0}^{\eta}\left(Z_{CS}(\eta')-1\right)\frac{d\eta'}{\eta'}\nonumber\\
&+&\rho\int_{1st \, shell}d\vec{R}\,g(\eta,R)\nonumber\\
&\times&\left[\beta
V_{model}^{eff}(R,\alpha_{liq};\beta\gamma,L_e)-\beta
V_{HS}^{eff}(R,\alpha_{liq})\right].
\end{eqnarray}
Here the first term gives the entropic cost when all the nearest-neighbor pairs are bonded,
with $\Lambda$ denoting the thermal wavelength. The second
term is the excess free energy of the HS reference
system where the compressibility factor $Z$ is given by
Carnahan-Starling(CS) equation of state. The last term involves
the energetic contributions from the attraction due to bond stretching;
note the HS part is carefully deducted.

We use SCP theory to describe the free energy of glassy configurations \cite{density functional,f.e.functional}.
The expression we use for an individual glassy configuration is
\begin{eqnarray} \label{f_gl}
f_{gl}&\equiv&\frac{\beta A_{gl}}{N}=\rho\int_{1st\,
shell}d\vec{R}\,g(\rho,R)\beta V^{eff}_{model}(R,\alpha_{gl};\beta\gamma,L_e)\nonumber\\
&+&\{\frac{3}{2}\ln\left(\frac{\alpha_{gl}\Lambda^2}{\pi}\right)
-3\ln\left[erf(\sqrt{\alpha_{gl}}D)\right]\}-\delta f.
\end{eqnarray}
Here the first integral gives the ``on-site" free energy after
double averaging over thermal fluctuations and over topological
disorder. The second term (inside the curly bracket) comes from
$-(1/N)\ln\left[Z_p\cdot\left(\int_{|w_i|\leq D}d\vec{w}e^{-\alpha
w^2}\right)^N\right]$ which accounts for the effect of cell
constraint; $Z_p$ represents the momentum part of the partition
function, and we choose a cubic cell with side length
$D\equiv\rho^{-1/3}/2$ for convenience. Finally, $\delta f$ (taken to be $0.224$)
is a numerical correction to the entropy of HS crystal system near face-centered-cubic (fcc)
close packing obtained via SCP approximation, which then gives extremely accurate free energies near melting.

The pressure can be evaluated by numerically differentiating the
liquid free energy:
\begin{equation}
p=\rho^2\left(\frac{\partial}{\partial\rho}f_{liq}\right)_{T,N}k_B
T.
\end{equation}

Random first order transition theory identifies the
configurational entropy with the difference between the free
energy of the highly localized glass solution and the liquid, i.e.
\begin{equation}
\frac{S_c(\rho,\beta)}{Nk_B}=\Delta f=f_{gl}-f_{liq}
\end{equation}

Another interesting quantity that can be evaluated within the SCP theory
is the number of force-bearing bonds, i.e.,\,those that have a
length exceeding the elasticity onset $L_e$. In three dimensions, the
structure-dependent probability for a single bond to be
elastically stretched (or equivalently, the fraction of stretched
bonds, in the mean field context) is given by
\begin{equation}
q_3(\alpha,R)=\int_{|\vec{r}|>L_e}d\vec{r}\,\left(\frac{\alpha}{\pi}\right)^{3/2}e^{-\alpha(\vec{r}-\vec{R})^2},
\end{equation}
or
\begin{equation}
q_3(\alpha,R)=\sqrt{\frac{\alpha}{\pi}}\frac{1}{R}\int_{L_e}^{\infty}dr\,r\left[e^{-\alpha(r-R)^2}-e^{-\alpha(r+R)^2}\right].
\end{equation}
The number of stretched bonds can then be obtained by averaging
over non-vibrational disorders, i.e.\ configurational degrees of
freedom:
\begin{equation}
q_3(\alpha,\rho)=\rho\int_{1st \,
shell}d\vec{R}\,g(\rho,R)q_3(\alpha,R).
\end{equation}
This double integral accounts both for the contributions from the
fluctuation of individual bonds (small $\alpha$ indicates strong
fluctuation) and from fluctuations in the underlying topological structure (short
$L_e$ enlarges the radial range that contains stretched ``fiducial
bonds").

\section{{Numerical investigations}}

\subsection{Localization strength}
The fundamental quantity that characterizes the diverse mechanical
phases in our model is the localization strength, or the force
constant of the emergent Einstein oscillator, $\alpha$. We may plot $\alpha$ against
bead density $\rho$ and/or effective stiffness
$\beta\gamma$ of the F-actin. We measure lengths and energies in units of $d$ and $\beta$, then the corresponding
dimensionless quantities are taken to be $\rho^*\equiv\rho d^3=\rho$ and $\gamma^*\equiv\beta\gamma d^2=\beta\gamma$. We start with several representative one dimensional plots that come from vertical
slices of the two dimensional $\alpha$-surface, and we
first focus on the liquid-like solution which is absent in the pure HS system.

\begin{figure}[htb]
\centerline{\includegraphics[angle=0, scale=0.26]{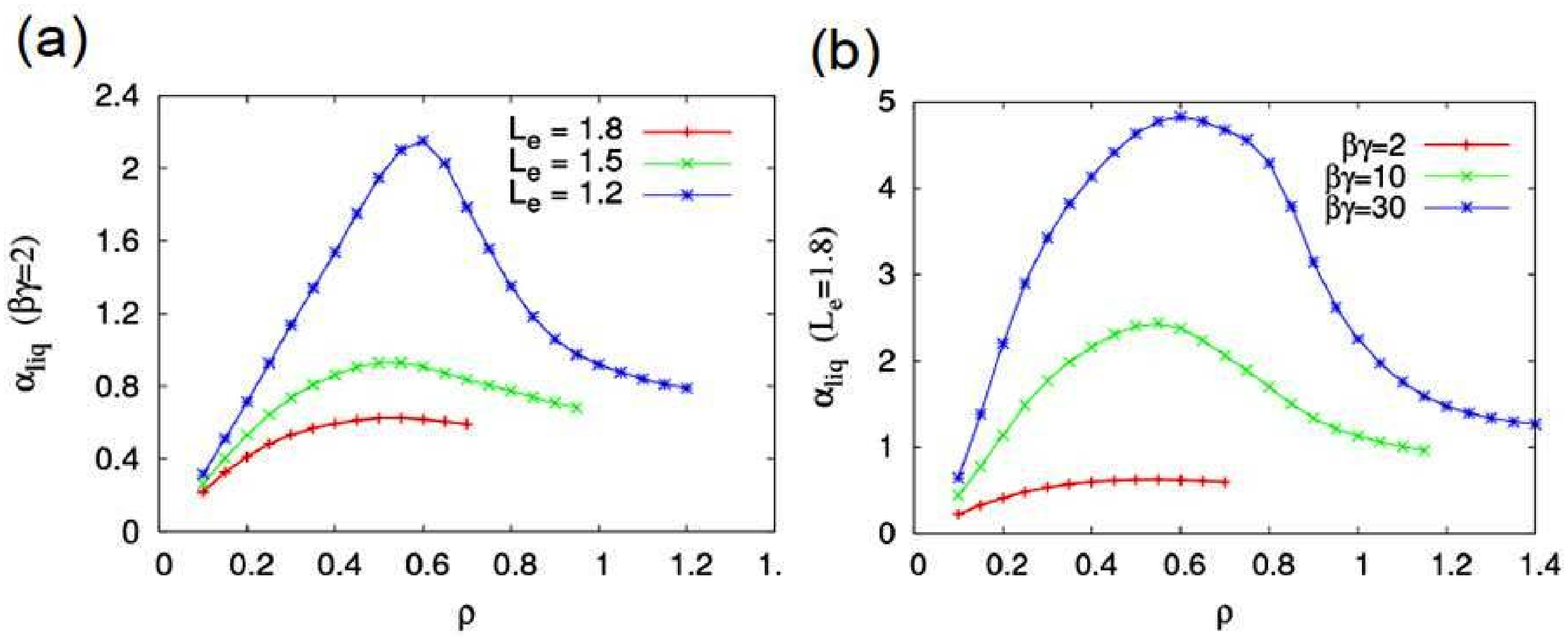}}
\caption{Liquidlike localization strength $\alpha_{liq}$ versus bead
density $\rho$. (a) For $\gamma^*=2$ with $L_e=1.2\
(blue), 1.5\ (green), 1.8\ (red)$; (b) for $L_e=1.8$ with $\gamma^*=2\ (red), 10\ (green), 30\ (blue)$.} \label{Lalpha_dune}
\end{figure}


Referring to Fig.~\ref{radialDist_1stShell}(a), we observe that the first shell
($R^*$) of the radial distribution shrinks from $1.95$ to $1.3$
as the bead density increases from $0.2$ to $1.2$. When
$L_e<R^*$, the separation ($R^*-L_e$) determines how many
nearest-neighbor fiducial sites are found beyond $L_e$ and
on-average have tense bonds; in this case, both the underlying topology
and thermal fluctuations contribute to the localization. On
the other hand, however, if $L_e>R^*$, all fiducial sites of
nearest neighbors fall inside the sphere of radius $L_e$ and on-average
result in buckled strings; in this case
fluctuations are the only source of stretching and thus of
localization. This information is also encoded in the
\textit{threshold density} $\rho_{th}$ beyond which a stable
$\alpha_{liq}$ solution no longer exists. In sum, longer onset
length leads to weaker overall localization as well as lower
threshold density, as shown in Fig.~\ref{Lalpha_dune}(a). A similar effect is produced by low
rigidity $\gamma$ as can be seen in Fig.~\ref{Lalpha_dune}(b); since smaller $\gamma$ indicates a broader
and shallower confining well in which particles are more loosely
tethered thus being less localized, and the corresponding liquid-like solution
becomes unstable at lower $\rho_{th}$.


\begin{figure}[htb]
\centerline{\includegraphics[angle=-90, scale=0.22]{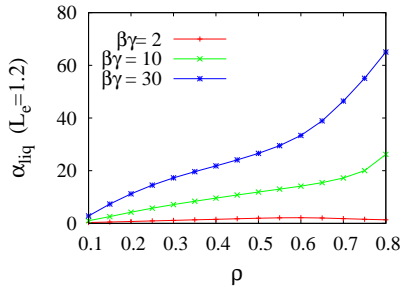}}
\caption{Liquidlike localization strength versus bead density for $L_e=1.2$ with $\beta\gamma=2, 10, 30$.
Liquid-like solution is still distinct at $\beta\gamma=2$, whereas a rapid crossover to glassy behavior occurs for $\beta\gamma=10, 30.$}
\label{Lalpha_L1.2}
\end{figure}

The compromise between the number of contributing neighbors and the degree of stretching
produces a non-monotonic density dependence of
$\alpha_{liq}$ at a given $L_e$, and the most efficient localization is achieved at
an intermediate density around $0.6$.
At $\rho<0.6$, the individual bonds become less likely to be stretched as the beads pack
more densely, whereas the total number of bonded neighbors increases much faster with increasing density,
the combined effect thus leads to a stronger localization as density grows.
On the other side when $\rho\geq0.6$, the number of bonded neighbors almost saturates,
while the thinner first shell of g(r) indicates less tightly stretching or deeper buckling as density increases, whereby the localization strength decays accordingly.

Note that this non-monotonic density dependence only occurs for a sufficiently long $L_e$ or in the low-$\beta\gamma$ regime for a short $L_e$. For sufficiently high $\beta\gamma$ liquid-like solutions are no longer distinct, instead a rapid crossover to high-$\alpha$ solutions is observed, as is exhibited in Fig.~\ref{Lalpha_L1.2}.

\subsection{Thermodynamics}
\subsubsection{$q_3$---role of localization}

\begin{figure}[h]
\centerline{\includegraphics[angle=0, scale=0.27]{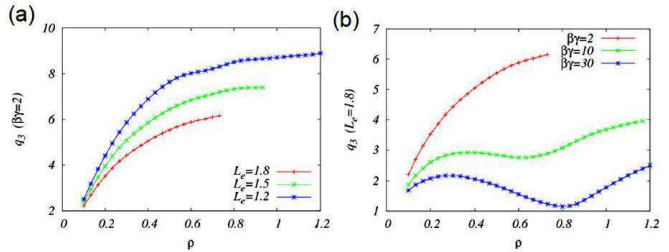}}
\caption{Average number of stretched bonds $q_3$ vs
bead density $\rho$. (a) $\gamma^*=2$ for $L_e=1.2,
1.5, 1.8$ (top to bottom); (b) $L_e=1.8$ for $\gamma^*=2,
10, 30$ (top to bottom).} \label{q3}
\end{figure}

Recall that $q_3$ explicitly counts the average number of tensely bonded neighbors.
Its dependence on elasticity comes from
the self-consistently determined localization strength $\alpha$
and the intrinsic cutoff $L_e$ for a bond to be stretched. For a given
$\rho$, more elastically-bonded neighbors become available as
$L_e$ drops (Fig.~\ref{q3}(a), bottom to top) and/or $\gamma^*$ decreases (Fig.~\ref{q3}(b), bottom to top).
For a moderate $\gamma^*$,
the number of stretched bonds increases with bead density for various $L_e$ as long as stable liquid-like solution exists (Fig.~\ref{q3}(a)); at high $\gamma^*$, however, close to the most localized regime for
individual beads, even though the coordination number still gently increases, the weak fluctuation (high $\alpha$) strongly suppresses bond stretching, thus leading to an intermediate decline of $q_3$ in its $\rho$-dependence, as seen in $\gamma^*=10, 30$ cases (Fig.~\ref{q3}(b)).

\subsubsection{free energy profile---consistency with SCP theory}

\begin{figure}[htb]
\centerline{\includegraphics[angle=-90, scale=0.27]{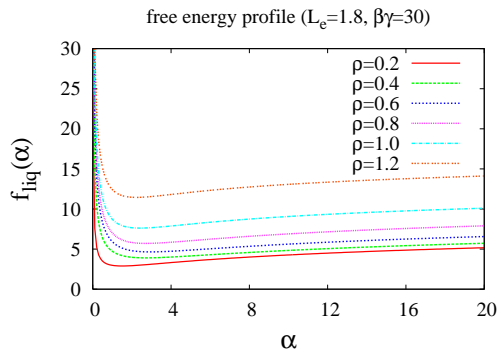}}
\caption{Free energy $f_{liq}$ versus the order parameter---localization strength $\alpha$ with
$L_e=1.8, \beta\gamma=30$ for $\rho=0.2$--$1.2$ (bottom to top).
As the bead density $\rho$ increases, the overall profile shifts upward
with the valley indicating the equilibrium solution $\alpha_{liq}$.}
\label{f.e._BETA30_L1.8}
\end{figure}

\begin{figure}[h]
\centerline{\includegraphics[angle=0, scale=0.28]{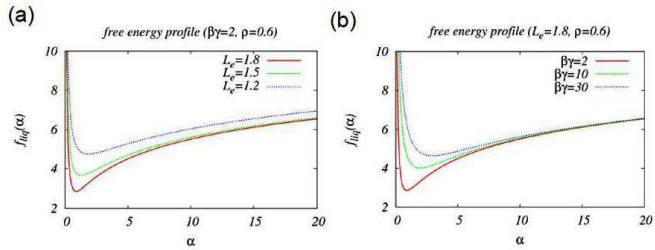}}
\caption{Free energy versus localization strength. (a) $\gamma^*=2$ for $L_e=1.2, 1.5, 1.8$ (top to bottom);
(b) $L_e=1.8$ for $\gamma^*=2, 10, 30$ (bottom to top).} \label{f.e.}
\end{figure}


As can be seen when we compare Fig.~\ref{f.e._BETA30_L1.8}
with the corresponding curves in Fig.~\ref{Lalpha_dune}(b),
the minima on the free energy profile $F(\alpha)$ for the
liquid-like state coincide with the corresponding $\alpha_{liq}$
solutions obtained via the SCP method. As the bead density increases, the
overall profile shifts upward. At a given density ($\rho=0.6$ here),
raising $L_e$ (Fig.~\ref{f.e.}(a))
or lowering $\gamma^*$ (Fig.~\ref{f.e.}(b)) causes the liquid-like state
to be increasingly favorable (deeper and sharper valley) with lower localization strength
(bottom of valley being shifted leftward). As expected, $\gamma^*$ only
markedly affects the low-$\alpha$ regime ($\alpha\leq10$), since when
particles are sufficiently localized (due to short $L_e$ or high
$\rho$), bond stiffness only plays a minor role in altering the
interaction strength. In contrast, varying $L_e$ not only
modulates the low-$\alpha$ regime, but causes a nearly uniform upshift of
the high-$\alpha$ portion of $F(\alpha)$; this observation is consistent with our previous statement that when thermal fluctuations are weak (high $\alpha$) topology dominates; shorter elasticity onset $L_e$ indicates more stretched bonds at a given density and thus enhanced interaction.

\subsubsection{pressure}

\begin{figure}[h]
\centerline{\includegraphics[angle=0, scale=0.25]{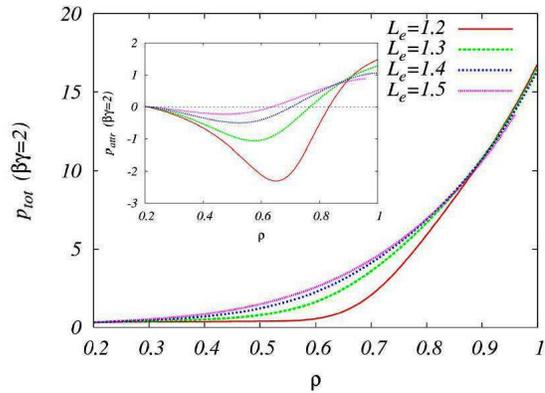}}
\caption{Pressure versus bead concentration for soft strings ($\beta\gamma=2$) with various onset lengths $L_e=1.2, 1.3, 1.4, 1.5$ (bottom to top). Main plot: total pressure; inset: pressure due to attraction.} \label{p_BETA2}
\end{figure}

We plot pressure versus bead density with $\gamma^*=2$ for a series of onset lengths in Fig.~\ref{p_BETA2}. At the low-density end, most of the bonds are stretched and the system tends to shrink, resulting in a very low pressure. At the opposite end, in a densely packed system HS repulsion takes over the major role; the steeply increasing pressure reflects the rising difficulty in rearrangement and the consequent soaring resistance to compression. In contrast with the nearly universal behavior at both extremal-ends, parameter-sensitive features emerge in the intermediate density regime, where shorter $L_e$ enhances the effective attraction and thus lowers the pressure of the system.
In particular, a non-monotonic behavior is observed to occur for pliable springs (low $\gamma^*$) with early onset (short $L_e$), as is exhibited for the case of $L_e=1.2$.

It is natural to expect the possibility of a negative pressure at a modest bead concentration in view of the tendency
of the network to shrink due to attraction. We plot in the inset of Fig.~\ref{p_BETA2} the contribution of
attraction ($p_{attr}$) to the total pressure ($p_{tot}$) for several onset lengths, and $p_{attr}$ does indeed
exhibit a negative valley at intermediate concentrations. It is clearly seen that a short onset length is necessary
for a notable contribution from attraction and thus for the emergence of this non-monotonic behavior of the total pressure.


On the other hand, the negative contribution owing to attraction is eventually overwhelmed by the positive part
due to a more rapid increase in repulsion as $\rho$ increases, and the total pressure ($p_{tot}$) remains positive over the whole density range. This behavior is originated from the specific form of the free energy functional
that we have used for the model system, where the correction due to attraction enters as a perturbation to the dominating repulsive part. This result might require modification to better accommodate the low-concentration regime
where attraction becomes important. Another reason for this behavior may be related to the $g(r)$ we have used for
the radial distribution which drops sharply when move away from the core; bonding effects might lead to
a fatter tail of $g(r)$ and an enhanced contribution from attraction. Nevertheless, we feel these insufficiencies
of the approximations should not modify the mechanical and thermodynamic properties qualitatively.



The thermodynamic considerations discussed above help with
understanding of the physics underlying the predicted phase
behavior in terms of the order parameter $\alpha$. We will
discuss these in detail next.

\subsection{State diagram}
\subsubsection{Overview}

In our model network, we assume permanent bond connections between
the nearest-neighbor pairs. The interplay between
these intrinsic constraints and the thermal fluctuations gives rise
to an inhomogeneous distribution of tensile forces throughout the
network. Such force heterogeneity exhibits as a (uniform)
non-vanishing tethering strength in our mean field context. This
explains the prediction that a completely ergodic fluid phase, which is
allowed for the pure hard-sphere system with short-range attraction and that has strictly
diffusive behavior for long times i.e. $\alpha\rightarrow0$ as
$t\rightarrow\infty$, never occurs for the equilibrium network
structure; the lack of utter freedom in locomotion leads to a
\textit{finite} $\alpha$ over the whole span of bead
concentrations.


There exist two characteristic densities in our model system:
the threshold density $\rho_{th}$ above which the homogeneous liquid-like solution is no longer stable, and the
critical density $\rho_{cr}$ which signals the emergence of a glassy
state. These densities help define the boundaries between diverse
mechanical phases:

$\bullet\quad\rho<\rho_{th},\rho_{cr}$: only $\alpha_{liq}$
exists, which describes the liquid-like loosely tethered phase;

$\bullet\quad\rho_{th}<\rho<\rho_{cr}$: the mean-field
$\alpha_{liq}$ is no longer stable, and the glassy state has not
yet occurred; bifurcation to a low-$\alpha$ solution may occur,
exhibiting a symmetry broken phase which we shall call the
``martensitic-like" (ML) phase;

$\bullet\quad\rho_{cr}<\rho<\rho_{th}$: both $\alpha_{liq}$ and
$\alpha_{gl}$ exist, depicting the transition state with
presumably coexisting phases that implies a macroscopic number of
configurational degrees of freedom for structural rearrangements.

$\bullet\quad\rho>\rho_{th},\rho_{cr}$: the repulsive-glass phase
dominates.

\subsubsection{Features of the various phases}


Our SCP calculation has found five distinct phases in our model system:
the liquid-like (LL) phase, the crossover (CO) phase, the repulsive-glass (RG) phase, the multiple-solution (MS) phase, and a martensitic-like (ML) phase.

$\Diamond$ the liquid-like (LL) phase and crossover (CO) phase


We choose two particular onset lengths based on their
position with respect to the boundary of the nearest-neighbor shell.

For $L_e=1.8$, this cutoff stands outside the nearest-neighbor shell for
$\rho:0.2$--$1.2$, which means that almost all the ``fiducial
bonds", i.e.\ bonds that connect fiducial sites of nearest
neighbors, are buckled, and become more floppy as
bead density grows; in this situation once the coordination
number saturates at $\rho\simeq0.6$, increasingly deeper buckling
results in a decrease in $\alpha_{liq}$ as $\rho$ increases.
Owing to the overall buckling of the bonds, higher bond stiffness and thus smaller fluctuations lead to a smaller probability for the bonds to tense up, and an even steeper drop in $\alpha_{liq}$ occurs as density increases (Fig.~\ref{Lalpha_dune}(b)). Therefore, the distinct liquid-like solutions persist over the whole $\gamma^*$-range of interest
(as long as $\rho\leq\rho_{th}$) for the case of a long onset length.

For the case of $L_e=1.2$, however, since this cutoff remains inside the nearest-neighbor shell
all the way through $\rho:0.2$--$1.2$, there always exists a fraction of fiducial bonds
being stretched beyond their relaxed length. In this situation the descending branch of
$\alpha_{liq}$ at intermediate densities only occurs for very soft springs, i.e.\,in the low-$\gamma^*$ regime.
Furthermore, taking advantage of the persistent fraction of stretched (fiducial) bonds, sufficient stiffness beyond a threshold value (marked by the phase boundary $\rho_{_{CO}}$) would help enhance localization, and facilitates a smooth crossover from elasticity-dependent liquid-like behavior to geometry-dominant glassy behavior (as seen in Fig.~\ref{Lalpha_L1.2}); we shall refer to the states showing such behavior as being in a ``crossover" (CO) phase.

$\Diamond$ the repulsive-glass (RG) phase


Current simulations performed on increasingly longer timescales make it possible to compare the
equilibrium properties based on the present theoretical predictions with simulation outcomes in the long-time limit.
To investigate the long-term fate of attractive glasses, simulations of glassy arrest in hard-core particles with short-range attraction \cite{bonding and caging simu} were performed over a waiting-time-independent window of up to $10^6$ MD units, which is a few orders of magnitude longer than those reached by previous experiments or simulations ($\sim 10^3$ MD units). They found that even if the short-range attraction generates a transient plateau
in the time-evolution of the calculated mean-squared displacement (i.e.\,inverse $\alpha$) owing to breaking and reforming of nearest-neighbor ``bonds", the long-time behavior of ``bonded" and
``nonbonded" repulsive glasses converges, suggesting that in the
long run particles are ultimately confined by their topological cage
of neighbors.


Our equilibrium calculation on a long-range-attractive HS system consistently shows that
as long as the (nearly-universal) critical density is reached, there would emerge the repulsive-glass behavior being almost independent of the attractive strength $\gamma^*$.
Such independence is found for various onset lengths.


$\Diamond$ the multiple-solution (MS) phase

The simultaneous presence of distinct $\alpha_{liq}$ and $\alpha_{gl}$ solutions,
according to the SCP analysis, implies a non-vanishing configurational entropy in this phase region.
This further signifies the availability of configurational degrees of freedom for structural rearrangement.
As shown in previous examples, a longer onset length ($L_e=1.8$) allows for distinct $\alpha_{liq}$ solutions over a larger $\beta\gamma$ range and thus allows the system to explore more energetically favorable configurations; on the other hand, the emergence of crossover behavior at moderate $\beta\gamma$ and the consequent shrinkage of the MS regime induced by early onset of elasticity ($L_e=1.2$) suggests a rapid loss of configurational entropy as bonds become stiffer.


$\Diamond$ the martensitic-like (ML) phase

As verified by our SCP calculation, a completely ergodic fluid phase, presented by a sticky HS system with weak attraction at a low density, does not occur for an elastically bonded HS system, so that $\alpha$ is always finite.
Yet finite localization strength does not ensure a homogeneous structure.
In our model system, with a large elasticity onset $L_e$ (compared to the lattice spacing) and a low bond stiffness $\gamma^*$, bifurcation in $\alpha_{liq}$ takes place before the emergence of glassy behavior (i.e.\ below $\rho_{cr}$) and may allow the existence of a spatial-symmetry-broken phase characterized by a stable pair of liquid-like solutions. The occurrence of bifurcation was also found in an earlier study by Shen and Wolynes with a pure ``cat's cradle" built on a regular lattice.
The possibility of self-generated spatial heterogeneity associated with such a mechanical instability makes it interesting to study what kind of structural phase such a destabilized system would actually develop into, and to quantify the associated conformational changes. The resultant mechanical/structural phase might be related to the orientational order often observed in F-actin networks of the cytoskeleton.

\begin{figure}[h]
\centerline{\includegraphics[angle=0, scale=0.25]{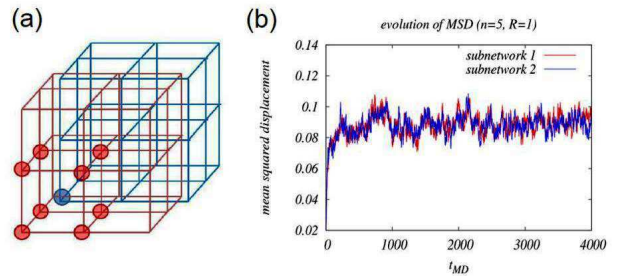}}
\caption{(a)\,An illustration of the subnetwork division. The bipartite nature of the original BCC lattice allows for separation of one SC subnetwork (red) from the other (blue). Each node on one subnetwork (blue sphere) interacts with its $8$ neighboring nodes (red spheres) on the other subnetwork.
(b)\,The MSD of both subnetworks ($n=5$) versus simulation time in MD units for a moderate elasticity onset to mean separation ratio $L_e/R=1$. $\Gamma=50, \beta=30$.}
\label{bcc_n5_MSD_R1}
\end{figure}

We performed explicit molecular dynamics (MD) simulations on a body-centered-cubic (BCC) network of $n^3$ unit cells
with pure nonlinear elasticity,
as described by the Hamiltonian $H=\frac{1}{2}\sum_{i}\sum_{<j>}\frac{1}{2}\gamma\left(|\overrightarrow{r_i}-\overrightarrow{r_j}|-L_e\right)^2
\Theta\left(|\overrightarrow{r_i}-\overrightarrow{r_j}|-L_e\right).
$
Here $i$,$j$ label the nodes and $\langle\cdot\cdot\cdot\rangle$ represents sum over nearest neighbors; $\gamma$ denotes bond stiffness and $L_e$ elasticity onset as before.
The bipartite nature of the BCC structure allows us to divide the original BCC network into two interpenetrating simple-cubic (SC) subnetworks (see an illustration of the subnetwork division in Fig.~\ref{bcc_n5_MSD_R1}(a))
, so that each node on one subnetwork interacts with its $8$ neighbors on the other subnetwork, and the neighbor list never changes.

The quality of the motion---whether it is oscillatory or monotonic---depends on the relative contribution of the inertial forces (that tend to produce oscillations) and the viscous forces (that tend to damp the oscillations out). It turns out that inertial forces are usually very small at the microscopic and molecular levels, so that the overdamped limit usually applies \cite{cytoskeleton}. We thus carried out the simulations with stochastic dynamics
in the overdamped limit as described by the Langevin equation
$d\vec r_i/dt=(1/\Gamma)\vec f_i(t)+\vec{\eta}_i(t)$.
Here $\vec f$ represents the deterministic force due to the nonlinear elastic interaction, and the force exerted by the fluid particles divides into two parts: the average viscous force $-\Gamma \vec v$ and a random force $\vec\zeta(t)\equiv\Gamma\vec\eta(t)$ whose time average is zero.
We assume a Gaussian white noise that satisfies $\langle\eta_i(t)\eta_j(t')\rangle=2D\delta_{ij}\delta(t-t')$.
As usual, $\Gamma$ and $D$ denote the drag coefficient and diffusion constant, respectively. Lengths are expressed in units of $L_e$.
We applied periodic boundary conditions.


We first show the evolution of the mean squared displacement (MSD) with respect to the \textit{initial} equilibrium positions of the nodes. We simulated a system of size $5^3$ with soft bonds ($b\equiv\beta\gamma=1$) at a low temperature ($\beta=30$).
When the elasticity onset $L_e$ is comparable with the initial mesh size, $R$, of both subnetworks, elastic stretching is immediately felt as soon as thermal buffeting displaces any of the nodes from their equilibrium locations, and therefore, the whole network just wiggles about the underlying BCC lattice, which defines the unique minimum of the system's energy landscape. As shown in Fig.~\ref{bcc_n5_MSD_R1}(b), in this case the two subnetworks act in a concerted manner with their mean squared displacements fluctuating almost ``in phase", suggesting the whole BCC network does not show symmetry breaking.

\begin{figure}[htb]
\centerline{\includegraphics[angle=0,scale=0.27]{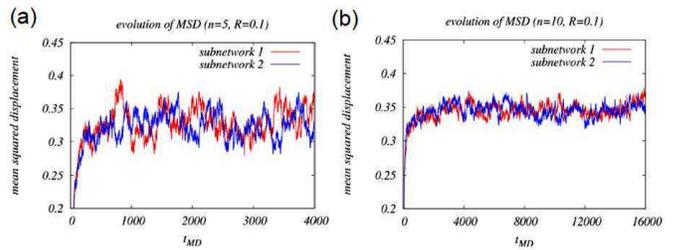}}
\caption{Effect of the system size on the behavior of MSD in the regime of bifurcation. Both: $L_e/R=10, \Gamma=50, \beta=30$. (a) $n=5$; (b) $n=10$.}
\label{n-dep}
\end{figure}

When the system parameters are modified to be in the regime of bifurcation by increasing the $L_e/R$ ratio to $10$ (other parameters unchanged), however, the MSD exhibits a much larger average amplitude (after a steady value is reached) and stronger fluctuations around it; furthermore, the two subnetworks fluctuate in a correlated manner but are almost completely out of phase with each other, see Fig.~\ref{n-dep}(a).
In this case, the initial network presents no elastic constraints on the nodes, thus the system is free to expand until it reaches a steady size where the mean separation between the bonded neighbors is comparable with the length for elasticity onset; the correlated fluctuations result from alternate distortions between the subnetworks within a system of moderate size as $5^3$. As we enlarge the system to $n=10$ (Fig.~\ref{n-dep}(b)), the steady amplitude of MSD maintains ($\sqrt{MSD}\sim 0.6$ stretching could occur under thermal driving), yet the fluctuations about the average become much weaker.

\begin{figure}[htb]
\centerline{\includegraphics[angle=0,scale=0.4]{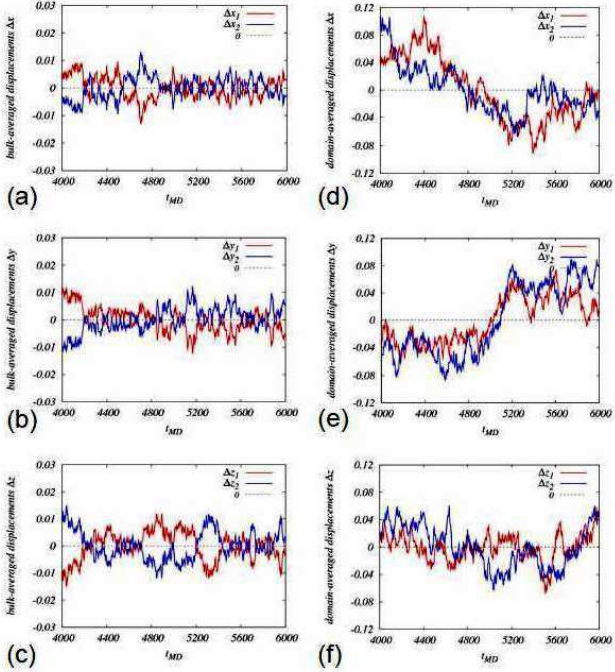}}
\caption{The bulk-averaged and domain-averaged displacement components for both subnetworks. $n=10, L_e/R=10, \Gamma=50, \beta=30$. (a)-(c)\,average over the whole simulated region; (d)-(f)\,average over one of the $8$ domains. (a)(d)\,x component; (b)(e)\,y component; (c)(f)\,z component.}
\label{avg_displacement_vector}
\end{figure}

The above observation raises the intriguing possibility of a ``martensitic-like" phase consisting of (frustrated) domains with (complementary) distortions. To examine further this possibility, we also queried vectorial information about the directions of motion (lost in MSD) by examining the evolution of displacement vectors. We divided the simulation cell evenly along each initial dimension to get $8$ domains each with the same number of nodes, and traced the subsequent motion within each domain.
We chose a time window $t_{MD}=4000$--$6000$ after the steady size was reached, and used the running time averaged position of the nodes as the reference with respect to which the displacements were defined. We present in Fig.~\ref{avg_displacement_vector} the bulk-averaged (left column) and domain-averaged (right column) displacement components. The evolution of the displacement vectors in these two cases shows several contrasting features. First, the maximal amplitude of the domain-averaged displacements is about one order of magnitude larger than the maximal amplitude of the bulk-averaged displacements. Second, the bulk-averaged displacements of the two subnetworks are exactly oriented in the opposition directions with essentially the same amplitude, whereas in individual domains two subnetworks move almost in the same direction yet with different amplitudes. These two contrasting features support the picture of localized distortions with different orientations in different domains.
In addition, the peak value of the displacement amplitude occurs at different times for different domains, which suggests ``frustration" between the domains.

To roughly estimate the domain size and to quantify the orientational correlation within the domains, we made equal-width shells centered at each node (the diameter of the outermost shell was taken to be equal to the box size of the simulated system), and computed two measures: (1) the average projection of director (i.e.\ the unit vector of the corresponding displacement) in each shell onto that of the central node, and then averaged over all possible central nodes. In mathematical terms, we monitor
$P_m(t)\equiv (1/N)\sum_{i=1}^N\hat{r}_i(t)\cdot\left(\frac{1}{N_m}\sum_{j_m=1}^{N_m}\hat{r}_{j_m}(t)\right)$,
where $N$ is the total number of nodes and $N_m$ the number of nodes within the $m^{th}$ shell of the central node;
$\hat{r}_i$ and $\hat{r}_{j_m}$ are instantaneous directors of the central and in-shell nodes, respectively;
(2) the average tensor product of directors constructed as $Q_m(t)\equiv(1/N)\sum_{i=1}^N\left(\hat{r}_i(t)\cdot\hat{y}\right)\left(\frac{1}{N_m}\sum_{j_m=1}^{N_m}
\left(\hat{y}\cdot\hat{r}_{j_m}(t)\right)\right)$, where $\hat{y}$ is the unit vector pointing from the instantaneous position of node $i$ to that of node $j_m$. This measure thus reflects the degree of alignment of two directors along the line connecting their positions.

\begin{figure}[htb]
\centerline{\includegraphics[angle=0, scale=0.28]{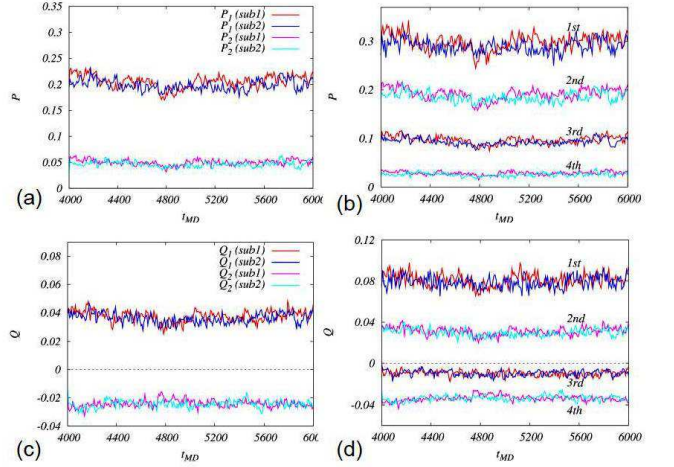}}
\caption{The evolution of two measures of the orientational correlation for both subnetworks. $n=10, L_e/R=10, \Gamma=50, \beta=30$. (upper) the average director projection P for $2$ shells\,(a) and $4$ shells\,(b); (lower) the average tensor product of directors Q for $2$ shells\,(c) and $4$ shells\,(d). In each panel, shown from top to bottom are values of the measure from the inner to the outer shell(s); red/magenta denotes subnetwork $1$, and blue/cyan denotes subnetwork $2$.}
\label{P_and_Q}
\end{figure}

We show in Fig.~\ref{P_and_Q} the evolution of these two measures for the cases of $2$ shells (left column) and $4$ shells (right column).
The average director projection $P$, in each shell, fluctuates about a steady value, which decays from around $0.2$ to $0.05$ as we go from the inner to the outer shell in the case of $2$ shells (panel (a)).
This spatial decay in average $P$ indicates that the domain size is roughly half of the box size, i.e., $5$ times the initial lattice spacing. As expected, the fluctuations in $P$ become weaker for outer shells due to the larger number of nodes to be averaged over; this trend is clearly exhibited in the case of $4$ shells (panel (b)).

In contrast to the positive steady value of $P$ in all shells, the average tensor product of directors $Q$ becomes negative for the outer shell(s) (lower panels of Fig.~\ref{P_and_Q}). Since $Q$ encodes the angular location of the in-shell node relative to the central node, a negative value of $Q$ implies the two directors point in the opposite directions along the line that connects them. A stronger opposite alignment would give a more negative $Q$ value. $Q$ thus serves as an indicator of the loss in orientational correlation. Also, negative $Q$s are smaller in amplitude than positive $Q$s.

\begin{figure}[htb]
\centerline{\includegraphics[angle=0, scale=0.2]{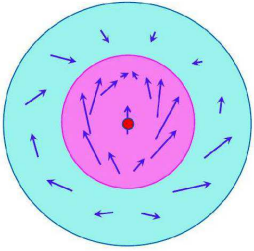}}
\caption{Average displacement vectors (purple arrows) at different angles and different distances (inner and outer shells) from the central node (red sphere). Shown for one cross section of the spherical region.}
\label{shell_sector}
\end{figure}

To make the situation more apparent, we visualized the ensemble-averaged displacement vectors at different angles and different distances from the central node for the case of $2$ shells (Fig.~\ref{shell_sector}).
We took the ensemble-averaged direction of motion of the central node as reference.
Within the first shell (colored in pink) the average displacement in any angular range gives a positive projection on the reference direction, in other words, the displacement field is anisotropic and the movements occur mainly along the same direction as that of the central node. The aligned movement is most significant near the ``equator".
The displacement vectors in the second shell (colored in blue), however, no longer exhibit a preferred orientation, instead they reorient considerably. Close to the ``south pole", the outward movements are almost perpendicular to the reference direction, whereas small inward movements occur near the ``north pole".
The spatial decay in orientational correlation is thus apparently observable and consistent with the features exhibited by our correlation measures.

In sum, in addition to the multiple solution regime which may be related to the saddle point solution pictured as ergodic droplets formed against a glassy background in a finite-range system, our mean-field level calculation also predicts a parameter regime where nonlinear-elasticity-induced spatial inhomogeneity is exhibited through a ``martensitic-like" phase with local oriented distortions.


\subsubsection{Typical examples}

We plot the 2D surfaces of $\alpha_{liq}$ and $\alpha_{gl}$ against
$\rho^*$(=$\rho$) and $\gamma^*$(=$\beta\gamma$) at a given $L_e$
to examine how these physical parameters modulate the phase boundaries.
The contour maps are also projected on the bottom as reference for
the upcoming state diagrams. We choose two particular values of the elasticity onset $L_e$ that characterize typical tense networks ($L_e=1.2$) and floppy networks ($L_e=1.8$).

\begin{figure}[htb]
\centerline{\includegraphics[angle=0, scale=0.27]{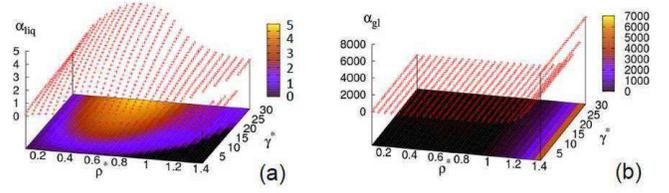}}
\caption{2D surface of localization strength over the parameter
space extended by bead density ($\rho^*$) and elastic
stiffness ($\gamma^*$) in the case of $L_e=1.8$ characterized by
mostly floppy bonds. Here $\rho^*$ runs through $0.1$ to $1.4$ and
$\gamma^*$ ranges from $1$ to $30$. We show the $\alpha$ surface
and its contour map for liquid-like\,(a) and glassy\,(b) solutions.
The color scheme indicates the relative measure of the $\alpha$ values;
the highest value within a given range is colored as bright
yellow and the lowest as black.} \label{2D_L1.8}
\end{figure}

We start with $L_e=1.8$ case. For this relatively long onset
length, liquid-like and glassy solutions are quite distinct over
the whole parameter space. As can be seen from Fig.~\ref{2D_L1.8}(a), the liquid-like localization strength takes on
a ``hump" shape along the $\rho$-axis peaking around $\rho=0.6$, and
elevates smoothly in the $\gamma^*$ direction. However, the stable
$\alpha_{liq}$ solution terminates at a sharp boundary defined
by $\rho_{th}(\gamma^*,L_e)$ beyond which the mean field
$\alpha_{liq}$ solution becomes destabilized. This instability region, located in the high-$\rho$ low-$\gamma^*$ corner, presents a quasi-triangular shape which indicates the increased need of stiffer bonds to stabilize the loosely arrested state as density increases. The ML phase may arise as a possible consequence of this (elastic-)nonlinearity-induced mechanical instability.
On the other hand, $\alpha_{gl}$ emerges at $\rho_{cr}\simeq1$. The values of the critical density and of $\alpha_{gl}$ are nearly independent of $\gamma^*$, as shown in Fig.~\ref{2D_L1.8}(b); such stiffness-independence arises from takeover of the dominant
role by HS repulsion in reconfiguring a densely packed system.

\begin{figure}[hbt]
\centerline{\includegraphics[angle=0, scale=0.45]{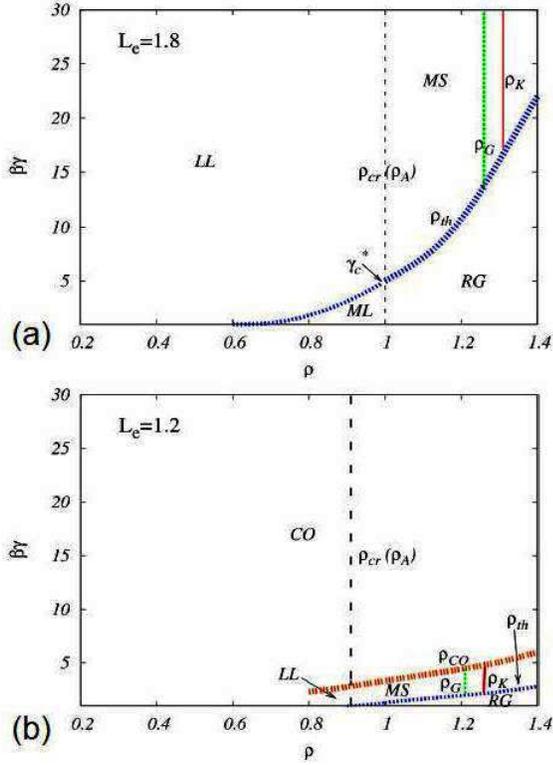}}
\caption{The state diagrams of a typical floppy network with $L_e=1.8$\,(a)
and a typical tense network with $L_e=1.2$\,(b).
These diagrams are constructed against the corresponding contour maps of liquid-like
and glassy solutions, summarizing all possible phases partitioned by
transition boundaries.
In the $L_e=1.8$ diagram, $\gamma^*_c$ marks the crossing point of $\rho_{th}$ curve (blue)
and $\rho_{cr}$ (or $\rho_A$) line (black dashed).
The ``mechanical" diagrams are further integrated with thermodynamic characteristics ---
laboratory glass transition density $\rho_{G}$ (green dotted line) and Kauzmann density $\rho_{K}$ (red full line).
The floppy network exhibits higher $\rho_G$ and $\rho_K$ than the tense network does.}
\label{diagram}
\end{figure}

The corresponding state diagram on $\rho$-$\gamma^*$ plane is
displayed in Fig.~\ref{diagram}(a). The
trajectories marked by $\rho_{th}$ (blue curve) and $\rho_{cr}$
(dashed line) unambiguously divide the state space into four distinct
phase regions: distinct liquid-like ($\rho<\rho_{th},\rho_{cr}$) and
glassy ($\rho>\rho_{th},\rho_{cr}$) phases locate at the opposite
corners diagonally, while the rest of the space naturally divides
into the ML and the MS phases depending on whether
$\gamma^*<\gamma^*_c$ ($\rho_{th}<\rho_{cr}$) or
$\gamma^*>\gamma^*_c$ ($\rho_{th}>\rho_{cr}$), respectively.
For a given $L_e$, $\rho_{th}$ increases with $\gamma^*$ whereas
$\rho_{cr}$ is topologically determined, consequently when
$\gamma^*<\gamma^*_c$ and $\rho_{th}<\rho<\rho_{cr}$, the region
of the ML phase narrows down as $\gamma^*$ increases due to decreasing
($\rho_{cr}-\rho_{th}$) until it disappears at $\gamma^*=\gamma^*_c$
($\rho_{th}=\rho_{cr}$); when $\gamma^*>\gamma^*_c$ and
$\rho_{cr}<\rho<\rho_{th}$, the MS phase takes over and
broadens as $\gamma^*$ rises because of growing
($\rho_{th}-\rho_{cr}$).

\begin{figure}[htb]
\centerline{\includegraphics[angle=0, scale=0.27]{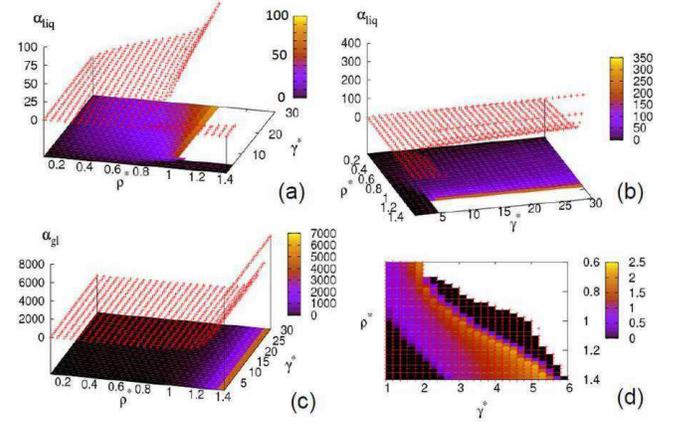}}
\caption{2D surface and contour map of the localization strength
for the case of $L_e=1.2$ that features a persistent fraction of tense bonds.
Upper panels: liquid-like solutions up to $\alpha_{liq}=100$\,(a) and $\alpha_{liq}=350$\,(b). Both the peeling-off stripe-shaped region for distinct liquid-like solutions and the smooth crossover to glassy behavior are explicitly displayed.
(d)\,Amplified contour map for the stripe-shaped region;
(c)\,glassy solutions over the whole parameter regime.}
\label{2D_L1.2}
\end{figure}

In addition to the four types of phases exhibited by the floppy network\,($L_e=1.8$), the state diagram for the tense network\,($L_e=1.2$) presents a novel phase boundary separating out a large region featuring a crossover behavior. When we compare the 2D surface for $\alpha_{liq}$ with that for $\alpha_{gl}$ over the whole parameter plane, we find that they almost coincide except for a stripe-shaped region in the low-$\gamma^*$ high-$\rho^*$ corner. When we zoom in on this region (contour map shown in Fig.~\ref{2D_L1.2}(d)) that peels off the smoothly ascending $\alpha$-surface (see upper panels of Fig.~\ref{2D_L1.2}), we observe that the diagram appears like a ``squeezed version" of that for the $L_e=1.8$ case; namely, $\alpha_{liq}$ also exhibits a non-monotonic density dependence, and $\rho_{th}$ locates the stability limit of such mean-field solution at each $\gamma^*$. As for the glassy state, however, $\alpha_{gl}$ proceeds with its $\gamma^*$-independent behavior as soon as the density exceeds $\rho_{cr}\simeq1$ (see Fig.~\ref{2D_L1.2}(c)). Therefore, similar to what was seen for the case of $L_e=1.8$, this stripe-shaped region presents LL, MS and RG phases. Noteworthy is that short $L_e$ dramatically reduces the region corresponding to the MS phase, indicating a rapid loss of the configurational degrees of freedom (upon bond stiffness increase). Moreover, the ML phase doesn't emerge for tense networks since the instability is avoided by early elasticity onset.

Beyond the upper boundary of this stripe-shaped phase region, the $\alpha$-surface gently mounts up toward
the high-$\gamma^*$ high-$\rho^*$ direction and smoothly crossovers from the elasticity sensitive behavior
to the glass-like behavior as $\rho_{cr}$ is approached; consequently, in this crossover regime $\rho_{cr}$
no longer marks a clear transition boundary.
The resultant state diagram is presented in Fig.~\ref{diagram}(b).

\subsubsection{Characteristic densities}

Due to the bonding constraints inherent in a network structure, the stabilized state
in our model always possesses a finite localization strength, i.e.\ $\alpha>0$.
Thus in this case the dynamical transition density $\rho_{cr}$ (or called $\rho_{A}$ as in literature),
rather than being the lowest density to give a non-zero $\alpha$ as occurred in pure
or sticky HS systems, is defined as the lowest density to trigger persistent
high-$\alpha$ solutions over the whole $\gamma^*$ range of interest. In our model system, $\rho_{cr}$ signals the
emergence of an extensive number of glassy metastable states, yet does not mark the
termination of bonding effect.

While the SCP theory alone allows us to find $\rho_A$, the ``Kauzmann density", $\rho_K$, at which the Helmholtz free energies of the liquid-like and glassy phases match and the configurational entropy ceases to be extensive also relies on the specific form of the free energy functionals we use for these two types of states. The ratio $\rho_A/\rho_K$ dimensionlessly characterizes the thermodynamic aspects. To connect to the kinetic laboratory glass transition, we note the laboratory transition is defined to occur when the viscosity reaches $10^4\,\mathrm{Poise}$. Random first order transition (RFOT) theory predicts this to be when the configurational entropy is about $1.0\,k_B$ per particle. To translate our thermodynamic results to the laboratory transition density, we will therefore mean by $\rho_G$ the
density where the liquid and glass free energies differ by $1.0\,k_BT$ per particle. Despite the universality of configurational entropy at laboratory transition experimentally confirmed in a wide variety of molecular glasses,
this universality must be examined further to see if it is valid in the cytoskeletal system which is an active biological material.
Here we just use this fiducial entropy to discuss the qualitative features of our model system.

\subsubsection{Possible transitions}

Visual inspection shows that the phase partition in the upper
portion ($\gamma^*>\gamma^*_c$) of $L_e=1.8$ state diagram
exhibits qualitatively identical behavior to that found near the bottom
of the stripe-shaped phase region in the diagram of $L_e=1.2$ case,
again indicating that high $\gamma^*$ and short $L_e$ are
comparably competent for making effectively more tense bonds. In
this regime, melting from RG via MS region to LL state is expected
as $\rho$ is lowered passing $\rho_{th}$ and $\rho_{cr}$ in
succession.
As for the lower section ($\gamma^*<\gamma^*_c$) of
$L_e=1.8$ diagram, upon increasing density, original homogeneous
LL phase becomes destabilized and develops into the proposed ML
phase where spatial heterogeneity develops, until finally the RG phase
takes the lead. At the crossing point, i.e. $\gamma^*=\gamma^*_c$,
RG melts into LL state without going via any intermediate phase.
In the case of $L_e=1.2$, as we go across the phase boundary $\rho_{_{CO}}$
by increasing density, the CO state would transform into the MS phase
whereby the distinction between the two types of arrested states
with different mechanisms of localization is recovered.

The effective bond stiffness can be varied by manipulating the crosslinking and/or bundling properties or by changing the temperature. Some general features can be extracted from the presented state diagrams,
that is high elastic stiffness tends to (1) stabilize the LL state and
(2) facilitate the crossover to glassy behavior.
The first effect is quite explicit in $L_e=1.8$ case:
as $\gamma^*$ increases, the elastic-nonlinearity-induced ML phase evolves into LL state
with a single stable $\alpha_{liq}$ when $\rho<\rho_{cr}$, whereas
purely RG state develops into the MS state when $\rho>\rho_{cr}$.
The second effect is clear in $L_e=1.2$ case; starting either
from distinct LL state or from the MS phase, the system would end up
with CO behavior as long as $\gamma^*$ transcends the phase
boundary $\gamma^*_{_{CO}}(\rho)$ (inversion of $\rho_{_{CO}}(\gamma^*)$).

Actually the parameter-modulated transformation of the phase behavior in terms
of the order parameter $\alpha$ can be directly detected in
$\alpha_{tagged}$ versus $\alpha_{neighbor}$ plots, which explicitly
show the emergence and disappearance of, as well as transitions among,
various fixed points under the parameter control. In other words, the
phase boundaries essentially indicate switching between different
fixed point structures of the self-consistent equations.
For example, the $\rho_{_{CO}}$ boundary marks the disappearance of the
lowest-$\alpha$ fixed point: in LL$\rightarrow$CO case, modest discontinuity
in $\alpha$ value arises from stability shift to a newly established fixed point
in proximity; while a considerable jump in $\alpha$ value is observed in
MS$\rightarrow$CO case, since no intermediate fixed point develops, the
high-$\alpha$ fixed point becomes the only stable attractor. Across the
$\gamma^*_{th}(\rho)$ (inversion of $\rho_{th}(\gamma^*)$) boundary from below, the initially bifurcation-generating
unstable fixed point becomes stabilized by enhanced stiffness.

In addition to the information obtained from the mechanical stability, the integrated
thermodynamic characteristics are more informative of the glassy aspects.
In the direction of vitrification,
kinetic laboratory glass transition is expected at $\rho_G$ in view of the landscape-dominated transport mechanism
triggered at $\rho_A$ and the presence of extensively many possible frozen-in states \cite{glass transition1,glass transition2,glass transition3}.
Whereas the Kauzmann density $\rho_K$, at which the configurational entropy ceases to be extensive
and the glassy configurations are no longer metastable, indicates a thermodynamic transition that
ultimately may underlie the kinetic arrest. Though hard to achieve on practical timescale, $\rho_{K}$ does
provide a mean-field estimate of how dense the liquid can be below which
a glass transition would be forced to intervene to avoid the entropy crisis \cite{entropy crisis}.


\subsubsection{Bonded-fraction dependence of mechanical and thermodynamical properties}

So far, we have implicitly assumed that all the nearest-neighbor pairs are bonded,
i.e.\,the network is fully connected and there are no free beads at all. In biological fact,
however, apart from the fibrous cytoskeletal network, there also exists a colloidal suspension of protein molecules (including detached ABPs, recycled actin monomers, etc.) that contributes equally, if not more, to the crowding interior of a cell, and thus to the excluded volume effect.
We mimic such a suspension of molecules simply as a collection of free beads, in which is immersed the nonlinear elastic fiber network which is anchored on the bonded beads. In our mean-field context, the fraction of bonded beads against the free ones is equivalent to the probability for a nearest-neighbor pair to be bonded. We assign an independent parameter $P_b\in(0,1]$ to indicate this bonded fraction or network connectivity, and assume $P_b$ to be independent of the overall bead density and the effective bond stiffness to purify its influence.

The self-consistent equation to determine $\alpha$ and the expressions of $f_{liq}$ and $f_{gl}$ are modified accordingly:

\begin{widetext}
\begin{equation}
\alpha=\frac{\rho}{6}\int_{1st\,
shell}d^3\vec{R}\,g(\rho,R)\,\bigg\{P_b\,Tr\!\left[\nabla\nabla\beta
V^{eff}_{model}(R,\alpha;\beta\gamma,L_e)\right]
+(1-P_b)\,Tr\!\left[\nabla\nabla\beta
V^{eff}_{HS}(R,\alpha)\right]\bigg\},
\end{equation}


\begin{eqnarray}
f_{liq}&=&P_b\left[\frac{3}{2}\ln\left(\frac{\alpha_{liq}\Lambda^2}{\pi e}\right)-1\right]+(1-P_b)\,\left(\ln\rho\Lambda^3-1\right)
+\int_{0}^{\eta}\left(Z_{CS}(\eta')-1\right)\frac{d\eta'}{\eta'}\nonumber\\
&+&P_b\,\rho\int_{1st \, shell}d^3\vec{R}\,g(\eta,R)\left[\beta
V_{model}^{eff}(R,\alpha_{liq};\beta\gamma,L_e)
-\beta V_{HS}^{eff}(R,\alpha_{liq})\right],
\end{eqnarray}

\begin{eqnarray}
f_{gl}&=&P_b\,\rho\int_{1st\,shell}d^3\vec{R}\,g(\rho,R)\beta V^{eff}_{model}(R,\alpha_{gl};\beta\gamma,L_e)
+(1-P_b)\,\rho\int_{1st\,shell}d^3\vec{R}\,g(\rho,R)\beta V^{eff}_{HS}(R,\alpha_{gl})\nonumber\\
&+&\bigg\{\frac{3}{2}\ln\left(\frac{\alpha_{gl}\Lambda^2}{\pi}\right)
-3\ln\left[erf(\sqrt{\alpha_{gl}}D)\right]\bigg\}-\delta f.
\end{eqnarray}
\end{widetext}

In equations for $\alpha$ and $f_{gl}$, $P_b$ and ($1-P_b$) lead the effective potential between bonded and non-bonded pairs, respectively; and the influence due to bonded fraction change on the cell constraint term (i.e.\,the second order term in the effective potential expansion) is contained in the self-consistently determined $\alpha_{gl}$. As for $f_{liq}$, the bonded fraction not only modifies the bonding correction to the HS interaction, it also separates the bonded from non-bonded contributions to the entropy cost: for free particles $\ln\rho\Lambda^3-1$ should suffice to describe the density dependence of the entropy cost, whereas we use $\frac{3}{2}\ln\left(\alpha_{liq}\Lambda^2/\pi e\right)-1$ for the bonded beads. Notice the fact that as bonds melt (i.e.\,$P_b$ decreases), increasing translational symmetry would lower the entropy cost to localize the density waves. We shall show that the logarithmic dependence on $\alpha_{liq}$ used here could at least qualitatively incorporate this feature. Moreover, the bonding entropy due to various choices of bonded pairs among nearest neighbors is not explicitly included, since only the difference between $f_{gl}$ and $f_{liq}$ matters for current purposes. It is easily seen that as $P_b\rightarrow1$ our earlier expressions for a fully connected network (Eqs.~(\ref{alpha})(\ref{f_liq})(\ref{f_gl})) are recovered.

\begin{figure}[hbt]
\centerline{\includegraphics[angle=0, scale=0.45]{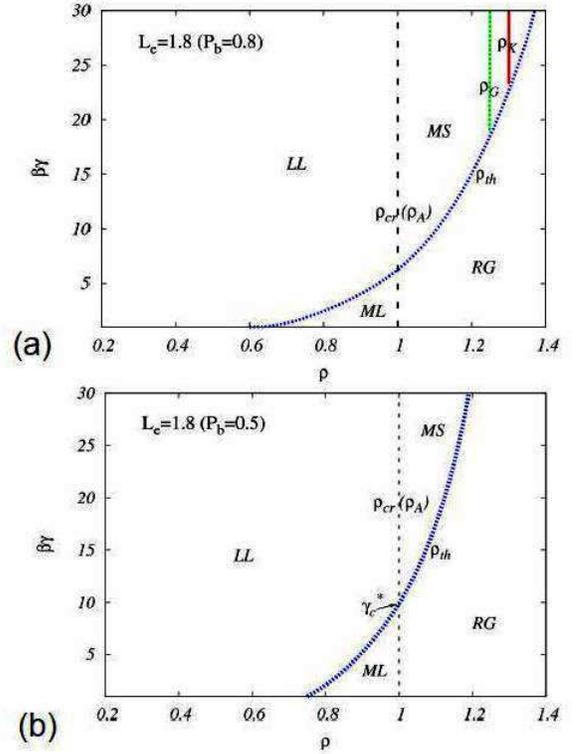}}
\caption{The state diagrams of a typical floppy network ($L_e=1.8$) with different bonded fraction.
$\rho_K$ (red) and $\rho_G$ (green) are defined as former.
(a)\,$P_b=0.8$; (b)\,$P_b=0.5$: transition densities are absent due to the limited $\beta\gamma$ range shown here.}
\label{diagram_L1.8_Pb}
\end{figure}

\begin{figure}[hbt]
\centerline{\includegraphics[angle=0, scale=0.45]{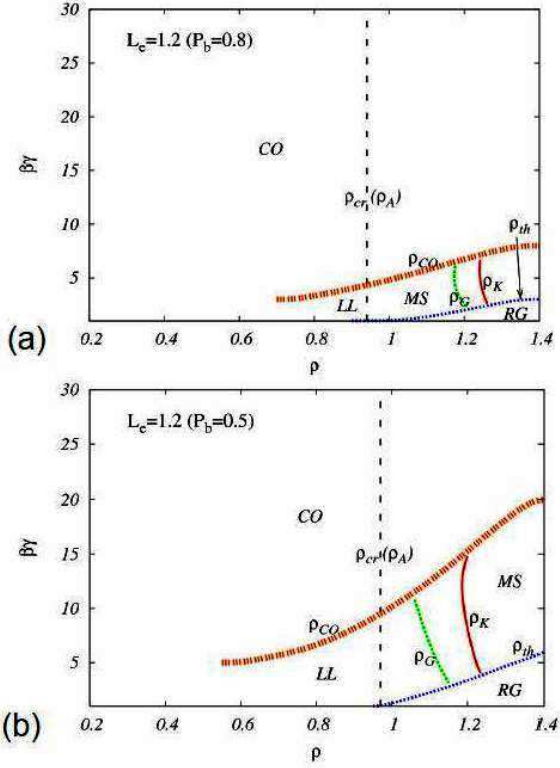}}
\caption{The state diagrams of a typical tense network ($L_e=1.2$) with different bonded fraction.
(a)\,$P_b=0.8$; (b)\,$P_b=0.5$. $\rho_G$ and $\rho_K$ become lower as the bonded fraction decreases.}
\label{diagram_L1.2_Pb}
\end{figure}

We show in Fig.~\ref{diagram_L1.8_Pb} and Fig.~\ref{diagram_L1.2_Pb} the state diagrams at $P_b=0.8$ and $P_b=0.5$ for both typical networks. The corresponding contour maps of $\alpha_{liq}$ and $\alpha_{gl}$ solutions (not shown here) indicate that the behavior of the liquid-like localization strength is not qualitatively affected by changing the bonded fraction, and the glassy solutions are almost quantitatively intact. Yet the transition boundaries are significantly shifted as the bonded fraction varies. For the floppy network\,(Fig.~\ref{diagram_L1.8_Pb}), as $P_b$ decreases, the slower increase of $\rho_{th}$ with $\gamma^*$ (which implies a lower destabilization density for the liquid-like solution) yields a shrinking MS region and enlarged RG and possibly ML regions. This behavior can be understood as arising from the fact that weaker network connectivity makes the stabilization via ``bond trapping" less efficient. In the tense network\,(Fig.~\ref{diagram_L1.2_Pb}), as the bonded fraction drops, both $\rho_{th}$ and $\rho_{_{CO}}$ boundaries shift upward resulting in an extension of both MS and RG regions into a higher $\gamma^*$ regime. This observation implies a higher bond stiffness is needed to stabilize the liquid-like solutions so as to trigger the crossover to glassy behavior. As for the dynamical transition density $\rho_A$ (i.e.\,the critical density $\rho_{cr}$ analyzed in earlier sections), it modestly increases with lowering $P_b$ in the tense network (rises from $0.91$ to $0.97$ as $P_b$ drops from $1.0$ to $0.5$) while it remains constant ($\sim1$) in the floppy network; concomitantly $\alpha_A$ halves its value ($115\rightarrow55$) in the tense network while it stays the same ($\sim50$) for the floppy case.


We next examine the variation in thermodynamics due to the change of bonded fraction. In both networks, $\rho_G$ and $\rho_K$ are found to persist upon decrease in $P_b$ (with the order of $\rho_A<\rho_G<\rho_K$ maintained). In the tense network, $\rho_G$ and $\rho_K$ develop moderate $\gamma^*$ dependence and shift toward lower density as $P_b$ decreases, as shown in Fig.~\ref{diagram_L1.2_Pb}. In the floppy network, $\rho_G$ and $\rho_K$ emerge at considerably higher $\gamma^*$ as $P_b$ drops indicating greater difficulty in stabilizing LL motion, yet become insensitive to $\beta\gamma$-value thereafter, as seen in Fig.~\ref{diagram_L1.8_Pb}(a). When $P_b$ is further lowered to $0.5$ (i.e.\,network being half-connected) transition densities are absent due to the limited range of $\beta\gamma$ shown here (see Fig.~\ref{diagram_L1.8_Pb}(b)) and would reappear if we extend $\beta\gamma$ sufficiently.

In contrast, if we use $\ln\rho\Lambda^3-1$ for both the bonded and non-bonded contributions to the entropy cost, a dramatic change in $\rho_G$ and $\rho_K$ is found (not shown here): when $P_b=0.8$, in both networks, $\rho_K$ is barely above $\rho_A$ while $\rho_G$ is entirely skipped; if $P_b$ is further lowered to $0.5$, then $f_{liq}>f_{gl}$ for all $\rho\geq\rho_A$ in MS region, indicating a negative configurational entropy which is not physically meaningful. Actually in our model the highly-localized glassy motion is insensitive to the degree of network connectivity since $\big<\beta V^{eff}_{model}(\alpha_{gl}; \rho>\rho_A;\gamma^*,L_e)\big>\simeq\big<\beta V^{eff}_{HS}(\alpha_{gl};\rho>\rho_A)\big>$ and $\big<\nabla^2\beta V^{eff}_{model}(\alpha_{gl}; \rho>\rho_A;\gamma^*,L_e)\big>\simeq\big<\nabla^2\beta V^{eff}_{HS}(\alpha_{gl};\rho>\rho_A)\big>$, thus such a significant drop in transition densities results from an enhanced attractive interaction in the liquid-like phase due to stronger thermal fluctuations (smaller $\alpha_{liq}$) induced by reduced bond constraints (lower $P_b$). This energetic enhancement in $f_{liq}$ is balanced, partly, by the decrease in entropic cost (to localize density wave) when $\frac{3}{2}\ln(\alpha_{liq}\Lambda^2/\pi e)-1$ is used, thereby mitigating the bonded-fraction modulation upon the transition densities, and resulting in a persistent transition possibility over a large $P_b$ range.

In sum, the overall tendencies are clear: decreases in the bonded fraction lower the transition densities $\rho_G$ and $\rho_K$ implying that the system becomes less capable of reconfiguring (or easier to become glassy) upon weakening of the network connectivity; in this sense, the model nonlinear-elastic-bonded interaction encourages liquid-like motion and thereby facilitates more efficient structural rearrangements.



\section{{Conclusions}\label{Conclusion}}

In this paper, we have modeled the cytoskeleton as an amorphous network of rigidly
cross-linked nonlinear-elastic bonds that become tense beyond an
intrinsic onset length and buckle otherwise. We study the
equilibrium mechanical properties of the model system within the
framework established by the self-consistent phonon theory and the
free energy functional formulation.

We have obtained an initial understanding of the physical behavior
via the calculation of several representative thermodynamic
quantities, and by examining the state diagram of typical systems.
Diverse mechanical properties of a generic cytoskeleton can be
recognized by analyzing the featured phases and possible
transitions: the permanent network structure excludes a completely
ergodic fluid phase, whereas the nonlinearity in the elastic interaction induces spatial heterogeneity that exhibits through a ``martensitic-like" phase with domains of oriented distortions.
The probable coexistence of the liquid-like and glassy behavior implies the capability of making structural rearrangements with varying agility in response to mechanical stimuli. The effective bond stiffness tends to stabilize the liquid-like state and facilitates its crossover to glassy behavior, whereas the relative position of the elasticity onset with respect to
the nearest-neighbor shell dramatically modulates the transition boundaries;
the critical density may no longer mark a sharp transition in certain situations when crossover takes place.
In sum, the elasticity onset length determines all possible mechanical phases within a practical parameter range, while the bond stiffness decides which transitions occur upon the variation
of cross-link concentration (thus of the bead density).

We further investigated how the bonded fraction or network connectivity modulates the phase boundaries as well as the thermodynamic terms of the transition densities ($\rho_G$ and $\rho_K$). We found that decreasing the bonded fraction results in an upward shift of both $\rho_{th}$ and $\rho_{_{CO}}$ boundaries, indicating the need for a higher bond stiffness to compensate for any loss in connectivity, so as to stabilize the liquid-like motion and to trigger its crossover to glassy behavior; on the other hand, the expanded multiple-solution phase region allows for a large stiffness range with extensive configurational degrees of freedom. As for the thermodynamics, the characteristic densities show little dependence on bond stiffness in floppy networks; for tense networks, however, $\rho_{G}$ and $\rho_K$ become lower upon enhanced bond stiffness, suggesting decreasing configurational degrees of freedom at a certain density as bonds stiffen. Further, for a given density, possible glass transition takes place at a much lower stiffness in tense networks than in floppy ones, while for a certain bond stiffness, the model system becomes vitrified at a lower bead density as more bonds tense up.

As exhibited clearly in the tense network, the logarithmic dependence of the entropy cost (for bonded interaction in the liquid-like phase) on particle localization strength adopted in our free energy functional contains the feature that as more bonds form, the kinetic glass transition would occur at a higher density, suggesting that the nonlinear-elastic-bonded interaction might help resolve local/steric constraints and facilitate escape from topological trapping, resulting in a more dense packing when stuck finally. Fleshing out this conjecture may require coming to term with finite range consideration; we need to go beyond mean-field level and consider activated events among various metastable states (probably via ``droplet relaxation" in a mosaic structure \cite{entropic droplet}). In a biological sense, cells may prefer an interconnected structural skeleton, not only to maintain their architecture, but to realize more efficient structural rearrangements when necessary.
As argued by Ingber,
the tensed/prestressed hierarchical networks play a central role in producing a well-orchestrated multiscale mechanical response\cite{prestress}.

This work provides a general scheme to study macroscopic mechanical phases in terms of the stability to local mechanical environment, and our current equilibrium model sets up a test field for further incorporated features for a more realistic model, in particular, the motorization effect that makes the system active, far from equilibrium and makes it physically distinct from an ordinary polymer network. It will be interesting to study the interplay of bond constraints and force-environment-sensitive motors in maintaining the cell's architecture and modulating the transition behavior. We also plan to investigate the effect of spatial heterogeneity and visualize the structural rearrangements by combining analytic schemes with simulation techniques.

\section*{ACKNOWLEDGEMENTS}
This work was supported by a National Science Foundation-sponsored Center for Theoretical Biological Physics Grant PHY-0822283.

\end{document}